\documentclass[12pt]{article}
\usepackage{graphics}
\usepackage{epsfig}
\usepackage[cp866]{inputenc}
\newcommand{\beq}{\begin{equation}}
\newcommand{\eeq}{\end{equation}}
\newcommand{\bea}{\begin{eqnarray}}
\newcommand{\eea}{\end{eqnarray}}
\newcommand{\eps}{\varepsilon}
\newcommand{\Fs}{\mbox{\scriptsize F}}
\newcommand{\pr}{_{\perp}}
\newcommand{\bk}{\bf {k}}

\newcommand{\Vef}{V^{p}_{\mbox{\scriptsize eff}}}
\newcommand{\DF}{\Delta_{\mbox{\scriptsize F}}}
\newcommand{\VF}{{\mathcal V}^{\mbox{\scriptsize F}}_{\mbox{\scriptsize eff}}}

\begin{document}

\begin{center}
{\Large\bf
Solution of the microscopic gap equation for a slab of nuclear matter
with the Paris NN-potential.}
\end{center}
\vskip 1 cm

\centerline{S.~S.~Pankratov$^{a}$, M.~Baldo$^{b}$, U.~Lombardo$^{c,d}$,
E.E.~Saperstein$^{a}$ and M~.V.~Zverev$^{a}$}

\vskip 0.5 cm
\begin{small}
\centerline {$^{a}$ Kurchatov Institute, 123182, Moscow,
Russia }
\centerline{$^b$INFN, Sezione di Catania, 64 Via S.-Sofia, I-95123
Catania, Italy}
 \centerline{$^c$INFN-LNS, 44 Via S.-Sofia, I-95123
Catania, Italy}
\centerline{$^d$ 44 Via S.-Sofia, I-95123 Catania,
Italy}
\end{small}


\vskip 0.5 cm

\begin{abstract}
The gap equation in the $^1S_0$-channel is solved for a nuclear slab with
the separable form of the Paris potential.
The gap equation is considered in the model space in terms of the
effective pairing interaction which
is found in the complementary subspace.
The absolute value of
the gap  $\Delta$ turned out to be very sensitive to the cutoff
$K_{max}$ in the momentum space in the equation for the
effective interaction.
It is necessary to take  $K_{max}{=}160 \div 180\;$fm$^{-1}$
to guarantee 1\% accuracy for $\Delta$.
 The gap equation itself is solved directly, without any additional
approximations. The solution reveals the surface enhancement of the
gap $\Delta$ which was earlier found with an approximate
consideration.
A strong surface-volume interplay was found also implying a kind
of the proximity effect. The diagonal matrix elements of $\Delta$
turned out to be rather close to the empirical values for heavy
atomic nuclei.
\end{abstract}
\newpage
\section{Introduction}

The microscopic theory of nuclear pairing is yet far from being
complete. Some progress was achieved only recently. In a series of
papers (see \cite{Rep} and Refs. therein), the problem was
attacked within a two-step approach in which the full Hilbert
space of the problem $S$ is split into the model subspace $S_0$ and the
complementary one, $S'$. The gap equation should be solved in the
model space in terms of the effective pairing interaction (EPI)
$\Vef$ which obeys the Bethe-Goldstone-type equation in the
complementary subspace. The pairing effects could be neglected in
the latter provided the model space is sufficiently wide. The
simplest nuclear systems with one-dimensional (1D) inhomogeneity
were considered, i.e. semi-infinite nuclear matter and the nuclear
slab. These simplifications permitted to find the EPI
directly
without additional approximations for the separable representation
\cite{Par1,Par2}  of the Paris NN-potential \cite{Paris}.
To save the computing time, it was suggested also an approximate method
of finding the EPI, the so-called local-potential approximation (LPA),
which has very high accuracy \cite{Rep}.

At the
same time, the gap equation in \cite{Rep} was solved only
approximately, with the help of a generalization to
non-homogeneous systems of the method  by V.A.~Khodel, V.V.~Khodel and
J.W.~Clark
(KKC) \cite{KKC}. Here we report the result of the direct solution
of the gap equation for the nuclear slab system.

In recent quite relevant paper by F.~Barranco et al. \cite{milan},
the direct solution of the gap equation for a finite nucleus
($^{120}$Sn) was carried out for the realistic Argonne $v14$
NN-potential. It yielded the gap $\Delta$ which is about a half of
the experimental value. The lack of the gap was attributed in
\cite{milan} to the contribution of the low-lying surface
vibrations. Comparison of the results of our calculation with
those of Ref. \cite{milan} will show to what extent the
predictions of the microscopic calculation of $\Delta$ in finite
systems depend on the form of the realistic NN-potential and other
details. In particular, it is well known \cite{Bal} that the gap
value in infinite matter is rather sensitive to the value of the
effective mass $m^*$. It is illustrated with Fig.~1 for the case
of the Paris NN-potential. Qualitatively, such a strong dependence
of $\Delta$ could be explained by the simplest BCS formula for
$\Delta$ in the case of the momentum independent force:
\beq \Delta = \eps_0 \exp\left(- \frac {m}{m^*g}\right),
\label{BCS} \eeq where $g$ is the dimensionless effective pairing
interaction strength, and the energy coefficient $\eps_0$ depends
on the cutoff in the integral over momentum in the BCS gap
equation.

\begin{figure}[h!]\vspace{-2mm}
\includegraphics [height=100mm,width=120mm]{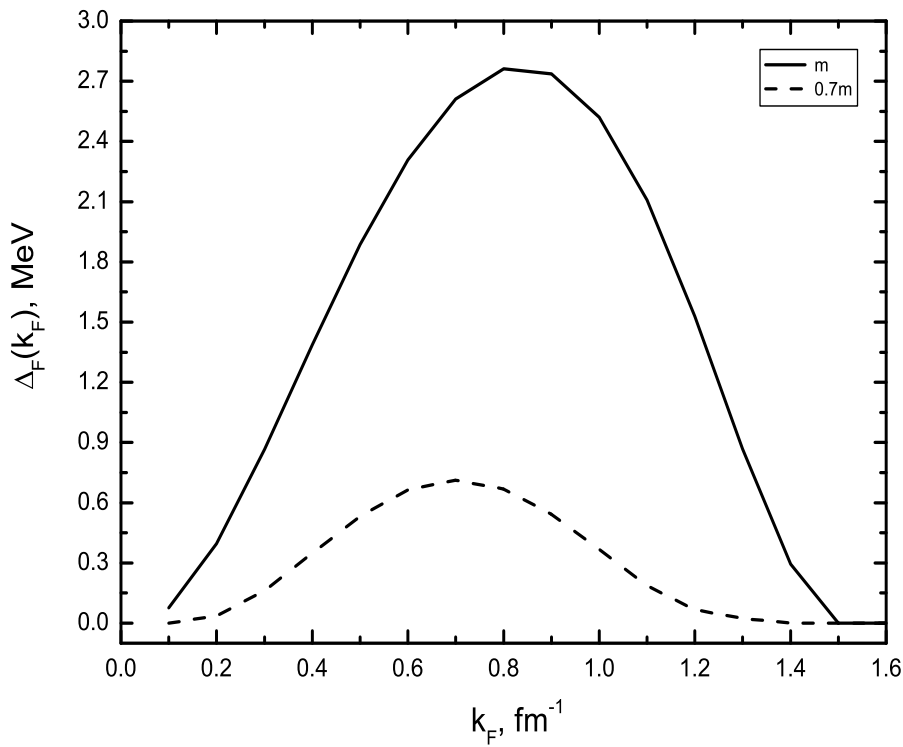}%
\vspace{0mm} \caption{ The gap $\Delta_{\Fs}(k_{\Fs})$ in nuclear
matter for Paris force. The solid line corresponds to the
effective mass $m^*=m$, the dashed one, to $m^*=0.7m$.}
\end{figure}

In \cite{milan} the coordinate dependent effective mass $m^*(r)$
was used of the Sly4 version \cite{Sly4} of the Skyrme force. Its
value changes from $m^*=0.7m$ inside a nucleus to $m^*=m$ outside.
We prefer to deal with the prescription of $m^*=m$ which is closer
to the ``experimental'' nuclear value. The latter was found in
Ref. \cite{KhS} from adjusting the single-particle spectra of
magic nuclei, in agreement with the older analysis by G.~E.~Brown
et al. \cite{Br}.

\section{Outline of the formalism}

Following to \cite{Rep}, we use
the many-body theory form of the gap equation \cite{AB,Schuck}
\beq \Delta(\eps) = - \int \frac {d\eps'}{2\pi i}{\cal V}
(\eps,\eps') A^s(E,\eps') \Delta(\eps'), \label{del} \eeq which
explicitly takes into consideration the particle-particle
propagator in the superfluid system,
\beq
 A^s(E,\eps) =  G(\frac {E}{2} + \eps) G^s (\frac {E}{2} - \eps).
\label{As}
\eeq
Here $G$ and $G^s$ stand for
the single-particle Green functions in normal and superfluid systems
correspondingly. Obvious space and spin variables in Eqs.~(\ref{del})
and (\ref{As}) are omitted. As usual, the product in Eq.~(\ref{del})
means the coordinate integration and spin summation.
The total two-particle energy is
$E=2\mu$, where $\mu$ is the chemical potential under consideration.

Within the Bethe-Brueckner approach, the irreducible
particle-particle interaction block ${\cal V}$ is approximated by
the free energy-independent $NN$-potential.  In this case, the gap
$\Delta$ is also energy-independent and the r.h.s. of
Eq.~(\ref{del}) can be integrated over $\eps'$ explicitly yielding
the well-known expression of the gap in terms of the anomalous
density matrix $\kappa$:
\beq \kappa({\bf r}_1,{\bf r}_2) =  \int   A^s(E;{\bf r}_1,{\bf
r}_2,{\bf r}_3,{\bf r}_4) \Delta({\bf r}_3,{\bf r}_4) d{\bf r}_3
d{\bf r}_4. \label{kap} \eeq

Here we use the same as above notation $A^s(E)$ for the integral
of Eq.~(\ref{As}) over $\eps$. In a symbolic notation, the
gap equation reads:
\beq
\Delta = - {\cal V} \kappa.
\label{delkap}
\eeq

When the solution of the gap equation in a finite system is carried out
in a direct way, the form (\ref{delkap}) is preferable as far as
the anomalous density matrix is expressed directly in terms of the
Bogolyubov $u$-, $v$-functions as follows:
\beq \kappa({\bf r}_1,{\bf r}_2) = \sum_i u_i({\bf r}_1) v_i({\bf
r}_2), \label{kapuv} \eeq where the summation is carried out over
the complete set of the Bogolyubov functions with energy
eigenvalues $E_i>0$.

The form (\ref{del}) of the gap equation is more convenient for
the two-step method of solution described in the Introduction.
After the splitting of the complete Hilbert space $S=S_0+S'$
described above, the two-particle propagator is represented as the
sum $A^s = A^s_0 + A'$. Here we already neglected  the
superfluid effects in the $S'$-subspace and omitted the
upperscript ``s'' in the second term.

The gap equation (\ref{del}) can be rewritten in the model subspace:
\beq \Delta = - {\Vef} A^s_0 \Delta, \label{del0} \eeq where
$\Vef$ obeys the following equation:
\beq {\Vef} = {\cal V} - {\cal V} A'{\Vef}. \label{EPI} \eeq

The model space $S_0$ is defined by a separation energy $E_0$ in
such a way that it involves all the two-particle states $(\lambda
,\lambda')$ with the single-particle energies $\eps_{\lambda},
\eps_{\lambda'} <E_0$. In this case, the complementary subspace
$S'$ involves such two-particle states for which one of the
energies $\eps_{\lambda},\eps_{\lambda'}$ or both of them are
greater than $E_0$.

The gap equation (\ref{del0}) in the model space can also be written
in the form
\beq \Delta = - \Vef \kappa_0, \label{del0kap} \eeq where
\beq \kappa_0 =  A^s_0 \Delta. \label{kap0} \eeq

However, a simple expression similar to Eq.(\ref{kapuv}) doesn't
exist for the anomalous density matrix $\kappa_0$ in the model
space. Indeed, let us substitute the usual pole expansions for
Green functions $G,G^s$ \cite{AB} for the integral over $\eps$ of
the expression similar to Eq.~(\ref{As}) for the model space propagator
$A^s_0$.
Let us then expand the $u$-, $v$-functions and the gap
$\Delta$ in the basis of the single-particle functions
$\phi_{\lambda}({\bf r})$ which makes the normal Green function $G$ diagonal,
 $G_{\lambda\lambda'}(\eps){=}G_{\lambda}(\eps) \delta_{\lambda\lambda'}$.
One gets:
\beq
u_i({\bf r}) = \sum_{\lambda} u_i^{\lambda} \phi_{\lambda}({\bf r}),
\label{ulam}
\eeq
\beq
v_i({\bf r}) = \sum_{\lambda} v_i^{\lambda} \phi_{\lambda}({\bf r}),
\label{vlam}
\eeq
\beq
\Delta ({\bf r}_1,{\bf r}_2) = \sum_{\lambda,\lambda'}
\Delta_{\lambda\lambda'} \phi_{\lambda}({\bf r}_1) \phi_{\lambda'}({\bf r}_2).
\label{delam}
\eeq

After simple operations, one finds
\bea \kappa_0 ({\bf r}_1,{\bf r}_2) =
\sum_{i,\lambda_1,\lambda_2,\lambda_3}^{\;\;\;\;\;\;\;\;\;\;(0)}
\left(\frac{n_{\lambda_1}\,v^{\lambda_2}_i\,v^{\lambda_3}_i}
{E_i-\eps_{\lambda_1}+\mu}+
\frac{(1-n_{\lambda_1})\,u^{\lambda_2}_i\,u^{\lambda_3}_i}
{E_i+\eps_{\lambda_1}-\mu}\right)
\nonumber\\
 \times\, \Delta_{\lambda_1\lambda_3}
\phi_{\lambda_1}({\bf r}_1) \phi_{\lambda_2}({\bf r}_2).
\label{kap0r}
\eea
Here $\eps_{\lambda}$ are the single-particle energies without pairing
and $E_i$, the ones with pairing taken into account.
The upperscript ``(0)'' in the sum means that all the states
$\lambda$  belong to the model space: $\eps_{\lambda} < E_0$.

  Let us now write down explicitly the Bogolyubov equations in
the ``$\lambda$-representation'':
\bea
(\eps_{\lambda}-\mu)\;u_i^{\lambda}+
\sum_{{\lambda'}}\Delta_{{\lambda}{\lambda'}}\;v_i^{{\lambda'}}
=E_i\,u_i^{\lambda}\,,
\nonumber \\
 \sum_{{\lambda'}}\Delta_{{\lambda}{\lambda'}}\;u_i^{{\lambda'}}
-(\eps_{\lambda}-\mu)\;v_i^{\lambda} =E_i\,v_i^{\lambda}\,,
\label{Bog3}\eea

With the use of Eqs.~(\ref{Bog3}), the sum in Eq.~(\ref{kap0r})
can be simplified to the following form: \beq \kappa_0({\bf
r}_1,{\bf r}_2) = \sum_{i,\lambda_1,\lambda_2}^{\;\;\;\;\;\;\;\;\;\;(0)}
\left(n_{\lambda_1}\,u_{i}^{\lambda_1}\,v_{i}^{\lambda_2}+
(1-n_{\lambda_1})\,u_{i}^{\lambda_2}\,v_{i}^{\lambda_1}\right)
\phi_{\lambda_1}({\bf r}_1) \phi_{\lambda_2}({\bf r}_2).
\label{kap01}
\eeq
It can be easily seen that in the limit of $E_0 \to \infty$,
when the set $(\lambda)$ is complete, the
expression (\ref{kap01}) coincides with Eq.~(\ref{kapuv}).

Following to \cite{Rep}, we use
the separable $3\times 3$ form \cite{Par1,Par2} of the
Paris potential \cite{Paris} in the $^1S_0$ channel,
\beq
{\cal V}({ k},{ k}^{\prime}) =
\sum_{ij} \lambda_{ij} g_i(k^2) g_j(k^{\prime 2}),
\label{Par}
\eeq
where the form factors $g_i(k^2),\; (i=1,2,3)$ are given in \cite{Rep}.

In this paper, we deal with a slab of nuclear matter
embedded in a potential well $U(x)$, where $x$ is the direction of the
1D-inhomogeneity. In this case, the momentum
${\bf k}_{\perp}$ in the $(y,z)$-plane (or ${\bf s}$-plane) is the integral
of motion.
The single-particle wave functions can be written as
\beq
\phi_{\lambda}({\bf r}) = e^{i{\bf k}_{\perp}{\bf s}} y_n(x),
\label{fi}
\eeq
and the Bogolyubov functions have the form:
\beq u_n({\bf r}) = e^{i{\bf k}\pr{\bf s}} \,u_i(k\pr,x)\, \eeq
\beq v_i({\bf r}) = e^{i{\bf k}\pr{\bf s}} \,v_i(k\pr,x)\,,
\label{uvx} \eeq

In this case, it is natural to use the
mixed representation, i.e., the coordinate representation for
the $x$-direction and the momentum one for the
 ${\bf s}$-plane. In this representation,
the separable NN-potential leads to a similar form
of $\Vef$ which, in the notation of \cite{Rep}, reads:
\beq
{\Vef}(k^2_{\perp},k^{\prime 2}_{\perp};x_1,x_2,x_3,x_4;E) =
\sum\limits_{ij}\Lambda_{ij}(X,X';E)
g_i(k^2_{\perp},x) g_j(k^{\prime 2}_{\perp},x').
\label{Vefij}
\eeq
Here the center-of-mass and relative coordinates
in the $x$-direction are introduced
($X=(x_1{+}x_2)/2$, $x=x_1{-}x_2$, etc.), and $g_i(k^2_{\perp},x)$
stand for the inverse Fourier transform of
the form factors $g_i(k^2_{\perp}+k^2_x)$ in the $x$-direction.
Their explicit form is given in \cite{Rep}.

The coefficients $\Lambda_{ij}$ obey the set of
integral equations:
\bea
\lefteqn{
\Lambda_{ij} (X,X';E) =\lambda_{ij} \delta(X{-}X') +
}
\nonumber \\
& &{} + \sum_{lm}\lambda_{il}\int dX_1\,B_{lm}(X,X_1;E)
\,\Lambda_{mj}(X_1,X';E), \eea where $B_{lm}$ are given by the
convolution integrals of the propagator $(-A')$ with two form
factors. Their explicit form is as follows:
\bea
B_{lm}(X,X';E) = { \sum_{nn'}}^{\prime}
           \int{ d^2{\bf k}_{\perp} \over {(2\pi)^2} }
{ (1 - N_{k_{\perp},n} - N_{k_{\perp},n'} )
\over {E -\eps_n-\eps_{n'}-k^2_{\perp}/m } }
\nonumber \\
\times
{g_{nn'}^l(k^2_{\perp},X)\,
g_{n'n}^{m*}(k^2_{\perp},X')},
\label{Blm}
\eea
\beq
   g_{n,n'}^l(k^2_{\perp},X) = \int dx
   \,g_l(k^2_{\perp},x)\,
 y_n(X{+}x/2)  y_{n'}(X{-}x/2).
\label{gnn}
\eeq
In Eq.~(\ref{Blm}) $N_{k_{\perp},n}{=}(0,1)$ are the occupation numbers
in the system without pairing.
The symbolic summation over ($nn'$) denotes
the summation over the discrete spectrum for negative energies and
the integration over $dp dp'/(2 \pi)^2$, for positive ones.
All the momentum integrals in (\ref{Blm}) are limited with a
cutoff, $p,p',k_{\pr} < K_{max}$.
The prime in the sum of Eq.~(\ref{Blm}) shows that the summation
is carried out over ($nn'$) which are not included  in the
model space of Eq.~(\ref{del0}) for $\Delta$. In the above short
notation, we have $(\lambda \lambda') \in S'$. Therefore, the energy
denominator in Eq.~(\ref{Blm}) always exceeds the difference
$(E_0-\mu) \gg \Delta$
what permits to neglect the pairing contribution to this propagator.

Analogously to Eq.~(\ref{Vefij}), we have
\beq \Delta(k_{\perp};x_1,x_2) = \sum\limits_{i}\Delta_i(X)
g_i(k^2_{\perp},x), \label{Deli} \eeq and
\beq \kappa_0(k_{\perp};x_1,x_2) = \sum\limits_{i}\kappa_0^i(X)
g_i(k^2_{\perp},x). \label{kapi} \eeq

The gap equation (\ref{del0}) is reduced to the set of three
one-dimensional integral equations,
\beq \Delta_i(X) = -
\sum_j\;\int\Lambda_{ij}(X,X';2\mu)\;\kappa_{0}^{j}(X')\;dX'.
\label{del0kapi} \eeq

For the 1D-geometry, the Bogolyubov functions can be expanded in the
basis of $y_n$-functions:
\beq u_i(k\pr,x) =
\sum_n u^n_i(k\pr)\,y_n(x)\,,\quad
v_i(k\pr,x) =
\sum_n v^n_i(k\pr)\,y_n(x)\,.\eeq

Then, for a fixed value of $k\pr$, the Bogolyubov equations have the form:
\bea  (\eps_n-\mu)\;u_i^n(k\pr)+
\sum_{n'}\Delta_{nn'}(k\pr)\;v_i^{n'}(k\pr)
=E_i\,u_i^n(k\pr)\,,
\nonumber \\
 \sum_{n'}\Delta_{nn'}(k\pr)\;u_i^{n'}(k\pr)
-(\eps_n-\mu)\;v_i^n(k\pr) =E_i\,v_i^n(k\pr)\,,
\label{Bog1}\eea
where
\beq\Delta_{nn'}(k\pr) =
\sum_{i}\;\int\Delta_{i}(X)\;g^{i}_{nn'}(k\pr^{2},X)\;dX.
\,\eeq

Then we have
\beq \kappa_0^j(X) =
\int\frac{d^{2}\bk\pr}{(2\pi)^{2}}\;\sum_{nn'}\,
\kappa_{0}^{nn'}(k\pr)\,g^{j}_{nn'}(k\pr^{2},X)\,,\eeq

where
\beq \kappa_{0}^{nn'}(k\pr) = \sum_i\left(N_{k\pr ,n} \,u_i^n(k\pr)\,
v_i^{n'}(k\pr)+ (1-N_{k\pr ,n})\,u_i^{n'}(k\pr)\,v_i^{n}(k\pr)\right).\eeq

The chemical potential $\mu$ in (\ref{Bog1}) is, as usual,
determined by fixing the average particle number $N$. In the
system under consideration, one should fix the number of particles
per a unit surface of the slab:
\beq \sigma = \int \frac{d^{2}{\bf
k}\pr}{(2\pi)^{2}}\,\sum_{in}\, \left(v_i^{n}(k\pr)\right)^{2} =
\mathrm{const}.
\label{detmu}
\eeq

The above set of equations is complete and can be solved by an
iteration procedure.

\section{Calculation results and discussion}

 In calculations, we used, just as in \cite{Rep},
the one-dimensional Saxon-Woods potential $U(x)$ symmetrical with
respect to the point $x=0$,
\beq
U(x) = \frac {U_0}{1 + \exp ((x-L)/d) + \exp (-(x+L)/d)},
\label{SW}
\eeq
with typical for finite nuclei values of the potential well depth
$U_0=50\;$MeV and diffuseness parameter of $d=0.65\;$fm.
For the main part of calculations the thickness parameter was chosen as
$L=8\;$fm to imitate heavy nuclei. The dependence of the gap $\Delta$ on the
value of $L$  is also examined.

It should be noted, that a simplification of the numerical procedure
in the slab system under consideration can be made using the
parity conservation which follows from the symmetry of the
Hamiltonian under the axis reflection $x \to -x$.
As a result, the eigenfunctions $y_n$ can be separated into
even, $y_n^+$, and odd, $y_n^-$, functions. Then
the two-particle propagators in Eqs.~(\ref{del0}),(\ref{EPI})
split into the sum
\beq
  A = A^+ + A^-
\label{Apm}
\eeq
of the even and odd components. The first one, $A^+$, originates
from the terms of the sum in Eq.~(\ref{Blm}) containing states
($\lambda,\lambda'$)
with same parity, and the second one, $A^-$, from those with opposite
parity. So long as  the NN-potential ${\cal V}$ does conserve
the parity, the propagators $A^+$ and $A^-$ do not mix in Eq.~(\ref{EPI}).
Therefore the separation similar to Eq.~(\ref{Apm}) holds true for the
correlation part of the EPI, $\delta \Vef{=}\Vef-V_0$.

As far as we deal with the
singlet $^1S_0$-pairing, the gap $\Delta$ is the even coordinate function.
Therefore only even components of the 2-particle propagators $A^s_0$,
$A'$ and of the correlation part $\delta \Vef$ enter Eqs. (\ref{del0}) and
(\ref{EPI}).

Let us now discuss several computational details. To simplify
account of the continuum, we used the discrete spectrum
representation method by putting the infinite wall at the distance
$R\ge 30\;$fm. Independence of results on the value of $R$ was
checked. The iteration procedure of solving the gap equation was
stopped when the maximum difference between values of
$\Delta_i(X)$ found in the consecutive iterations became less than
$10^{-5}\;$MeV. The typical number of the effective iterations was
40. At each iteration, the chemical potential $\mu$ entering the
Bogolyubov equations (\ref{Bog1}) was found anew from
Eq.~(\ref{detmu}).

By solving the gap equation, one finds a set of the components
$\Delta_i(X),\,i{=}1,2,3$ of the gap function and the ones,
$\kappa_i(X)$, of the anomalous density matrix. To present the results
in a more transparent way we introduce, following to \cite{Rep},
the ``Fermi averaged'' gap
\beq
\DF(X)=
\sum_i  \Delta_i(X)
g_i(k^2_{\Fs}(X)),
\label{Del_F}
\eeq
and density matrix,
\beq
\kappa_{\Fs}(X)=
\sum_i  \kappa_i(X)
g_i(k^2_{\Fs}(X)).
\label{kap_F}
\eeq
Here the local Fermi momentum is defined as follows:
$k^2_{\Fs}(X){=}2m(\mu{-}U(X))$ if  $2m(\mu{-}U(X)) > 0$,
and $k^2_{\Fs}(X){=}0$ otherwise.
For the Fermi averaged EPI, we have
\beq
{\VF}(X)=
\sum_{ij} \bar {\Lambda}_{ij}(X)
g_i(k^2_{\Fs}(X)) g_j(k^2_{\Fs}(X)),
\label{Vef_F}
\eeq
where the zero moments of the EPI components ${\Lambda}_{ij}(X_1,X_2)$
are defined in a standard form:
\beq
\bar{\Lambda}_{ij}(X) = \int dt \,\Lambda_{ij} (X{-}t/2,X{+}t/2).
\label{Lam_m0}
\eeq

  We found that the model space with the maximum energy  $E_0{=}20\;$MeV
is sufficiently wide  for convergence of the calculation procedure.
It is illustrated in Fig.~2.

\begin{figure}[h!]\vspace{-2mm}
\begin{center}
\includegraphics[height=80mm,width=100mm]{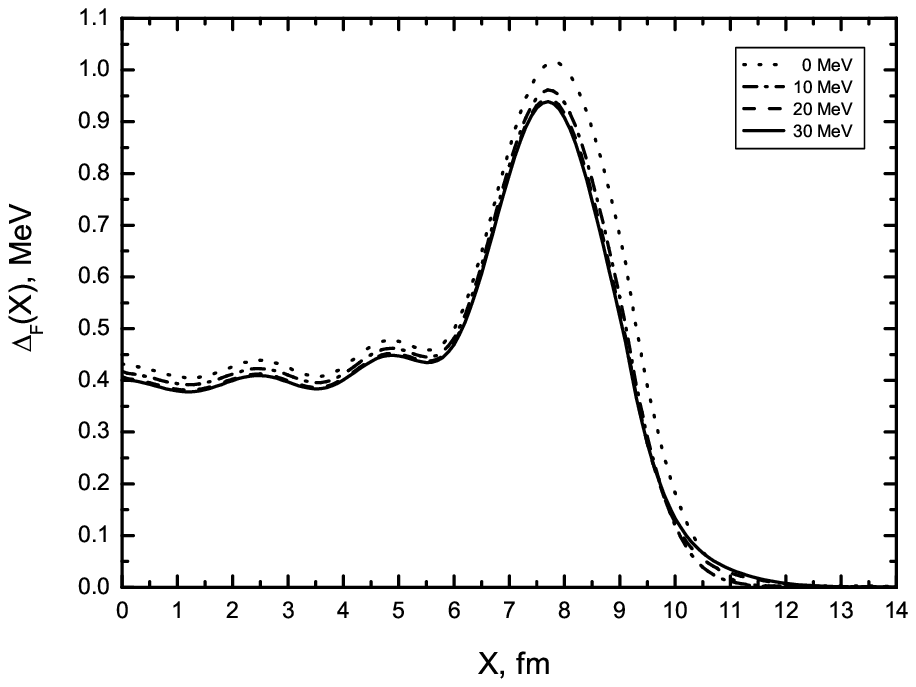}%
\vspace{0mm}
\end{center}
\caption{ The Fermi averaged gap $\DF(X)$ for different $E_{0}$.}
\end{figure}

It is seen that the function $\DF(X)$ changes
noticeably under variation from  $E_0{=}0$ to  $E_0{=}10\;$MeV and
only a little, from $E_0{=}10\;$MeV to $E_0{=}20\;$MeV. Finally,
it doesn't practically change
under variation from $E_0{=}20\;$MeV   to  $E_0{=}30\;$MeV.
The Fermi averaged EPI at different $E_0$ is displayed in Fig.~3.

\begin{figure}[h!]\vspace{-2mm}
\begin{center}
\includegraphics[height=80mm,width=100mm]{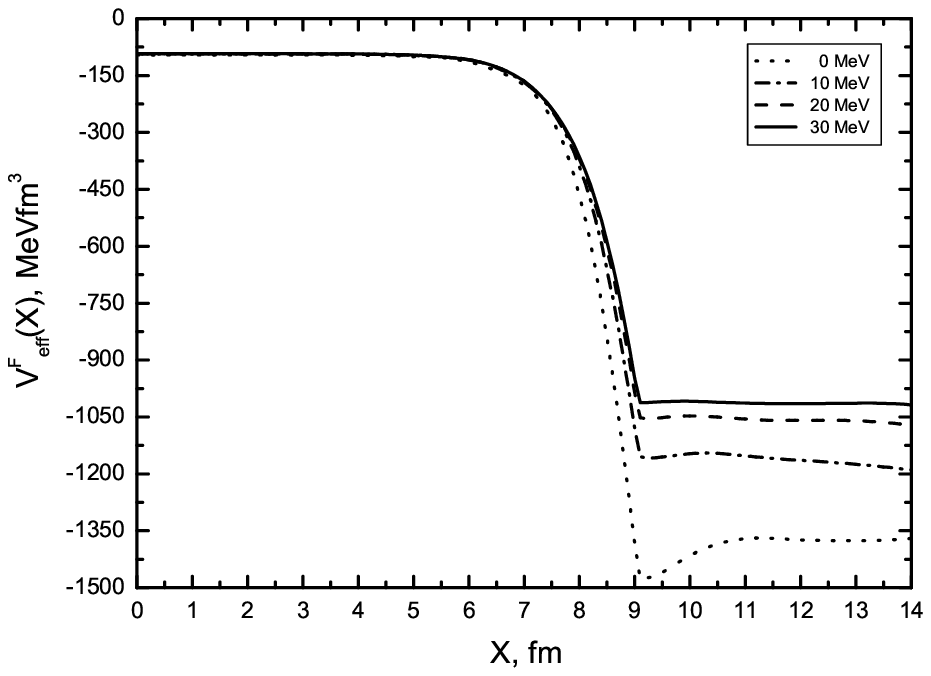}%
\vspace{0mm}
\end{center}
\hskip 1.5 cm
\begin{minipage}[c]{0.8\linewidth}
\caption{ The Fermi averaged EPI  $\VF(X)$ for different $E_{0}$.}
\end{minipage}
\end{figure}

It reveals a strong attraction outside the slab which changes
rapidly in the surface region to rather weak attraction inside the
slab. The asymptotic value of ${\VF}(X \to \infty)$ decreases with
growing of $E_0$. As $E_0 \to \infty$, the function ${\VF}(X)$
must tend to the analogous quantity ${\cal V}_0^{\Fs}$ calculated
for the free NN-interaction. The  asymptotic value of ${\cal
V}_0^{\Fs}$ is ${\cal V}_0^{\Fs}(X \to \infty){=}-720\;$MeV.

The situation with convergence of the integrals of Eq.~(\ref{Blm})
for the 2-particle propagator turned out to be more dramatic. A
very slow convergence of this integral was noted in \cite{BLSZ1}
where the problem of microscopic calculation of the EPI was
analyzed, but the gap equation itself was not solved. Such a big
value of the cutoff momentum as $K_{max}{=}60\;$fm$^{-1}$ was
chosen to guarantee a reasonable accuracy of the EPI. The main
reason  of this phenomenon is the hard-core character of the Paris
potential. Now we found that the gap $\Delta$ is much more
sensitive to the cutoff than the EPI, and such a huge value as
$K_{max}{=}160 \div 180\;$fm$^{-1}$ is necessary to obtain
$\Delta$ with 1\% accuracy. It can be seen from Fig.~4 for the
Fermi averaged gap $\DF(X)$ calculated for different values of
$K_{max}$. The corresponding results for the EPI are displayed in
Fig.~5.

\begin{figure}[h!]\vspace{-2mm}
\begin{center}
\includegraphics[height=80mm,width=100mm]{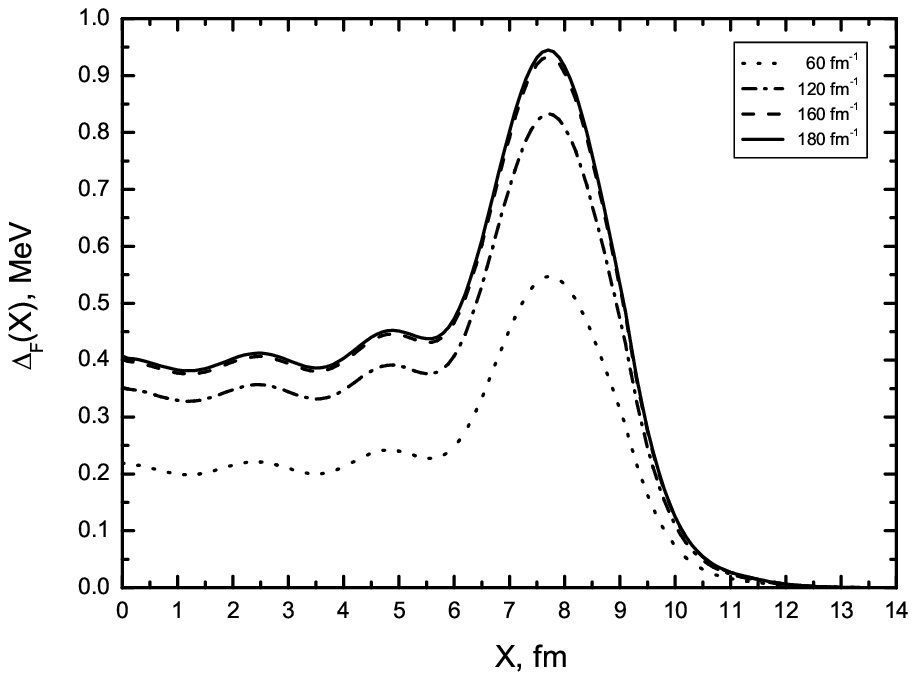}%
\vspace{0mm}
\end{center}
\hskip 1.5 cm
\begin{minipage}[c]{0.8\linewidth}
\caption{ The Fermi averaged gap $\DF(X)$ for different cut-off
momenta $K_{max}$. The model space corresponds to $E_{0}=20$ MeV.}
\end{minipage}
\end{figure}

\begin{figure}[h!]\vspace{-2mm}
\begin{center}
\includegraphics[height=80mm,width=100mm]{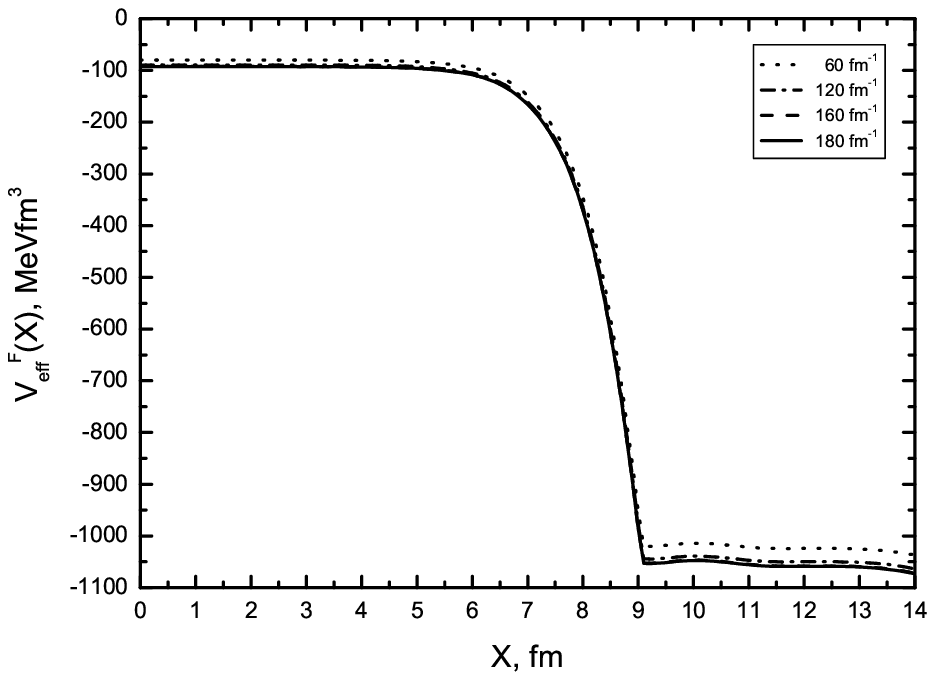}%
\vspace{0mm}
\end{center}
\hskip 1.5 cm
\begin{minipage}[c]{0.8\linewidth}
\caption{ The Fermi averaged EPI  $\VF(X)$ for different cut-off
momenta $K_{max}$. The model space corresponds to $E_{0}=20$ MeV.}
\end{minipage}
\end{figure}

\begin{figure}[h!]\vspace{-2mm}
\begin{center}
\includegraphics[height=80mm,width=100mm]{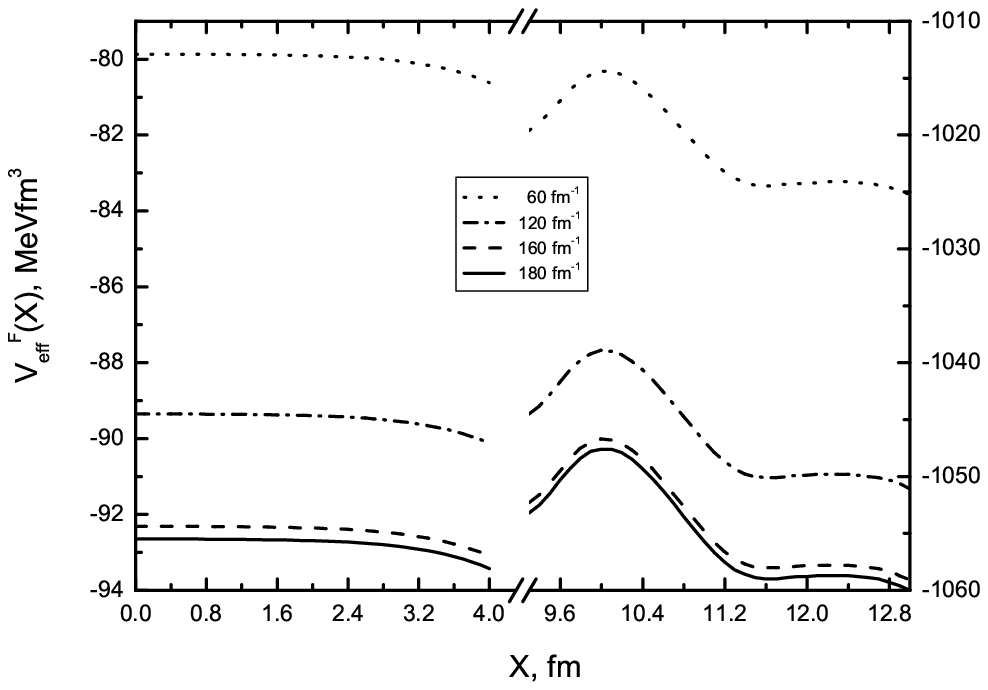}%
\vspace{0mm}
\end{center}
\hskip 1.5 cm
\begin{minipage}[c]{0.8\linewidth}
\caption{ Details of Fig.~5 in an enlarged scale. }
\end{minipage}
\end{figure}

To emphasize the difference between the curves, we displayed in
Fig.~6 the regions of small and large $X$ separately, in a
enlarged scale. It is seen that the variation of the EPI with
growing of $K_{max}$ from 60 fm$^{-1}$ to 180 fm$^{-1}$ looks
quite small, but it leads to a noticeable enhancement of $\DF(X)$.
In fact, 1\% variation of EPI results in 5\%, or even more, change
of $\Delta$. The  reason of such a high sensitivity of $\Delta$ to
a small variation of the EPI could be seen in the exponent
dependence of $\Delta$ on the inverse effective interaction
strength $1/g$ in the BCS formula (\ref{BCS}). In this
approximation, all depends on the value of $g$. Experience of
solving the gap equation  with Paris potential in infinite nuclear
matter shows that in the region of the maximum of $\Delta$ ($k_F
\simeq 0.8\;$fm$^{-1}$ ), where the effective $g$ is big, the
convergence is not so slow. In this case,
$K_{max}{=}20\;$fm$^{-1}$ is enough to reach  1\% accuracy in
$\Delta$. However, let us consider the density describing the
inner region of the slab under consideration,
$k_F{=}1.42\;$fm$^{-1}$, which corresponds to a small $\Delta$
(small $g$) situation. In this case, the convergence is very slow,
and for 1\% accuracy in $\Delta$ one needs in $K_{max} \simeq
200\;$fm$^{-1}$. Looking in Fig.~5, one could conclude that the
surface region with a big EPI should dominate in the gap equation
(\ref{del0kapi}). In this case, the sensitivity of $\Delta$ to
small variation of the EPI should not be  so high. However, the
comparison of Fig.~4 and Fig.~5 shows that the inner region with
small EPI (and high sensitivity of $\Delta$ to $g$) should
contribute to Eq.~(\ref{del0kapi}) significantly to explain this
strong influence of a variation of $K_{max}$ on the value of
$\Delta$.

To check this point, it is instructive to analyze the Fermi averaged
anomalous density matrix $\kappa_{\Fs}(X)$. It is displayed in Fig.~7
together with the EPI $\VF(X)$ and the product of $\VF(X)\,\kappa_{\Fs}(X)$.

\begin{figure}[h!]\vspace{-2mm}
\begin{center}
\includegraphics[height=80mm,width=100mm]{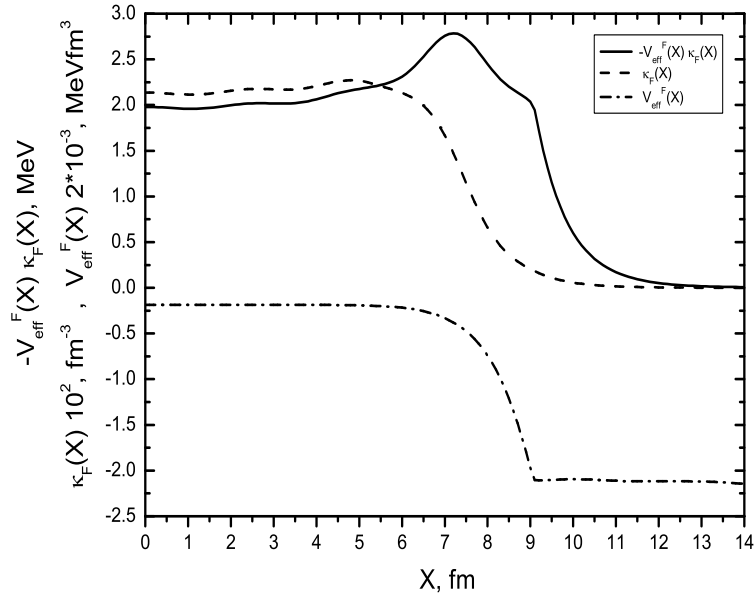}%
\vspace{0mm}
\end{center}
\hskip 1.5 cm
\begin{minipage}[c]{0.8\linewidth}
\caption{ The Fermi averaged anomalous density  $\kappa_{\Fs}(X)$,
EPI $\VF(X)$ and their product, in arbitrary units. }
\end{minipage}
\end{figure}

All the functions are drawn in arbitrary units, for the graphical
convenience. It is seen that the anomalous density matrix falls
rapidly down outside the slab. This decrease almost compensate the
surface increase of the EPI in Eq.~(\ref{del0kapi}). Indeed,
although the product function inside the slab is less than at the
surface, the difference is not significant.

To examine a relevance of our previous approximate calculations of
$\Delta$, it is worth to consider the so-called gap-shape
function which was found in \cite{Rep} with the help of a
generalization of the KKC method \cite{KKC} to non-homogeneous systems.
It was defined as follows:
\beq
\chi(x_1,x_2;k^2_{\perp}) =
\Delta(x_1,x_2;k^2_{\perp})/\DF(0).
\label{chi}
\eeq

Obviously, it can be represented as the sum
\beq
\chi(x_1,x_2;k^2_{\perp}) =
\sum\limits_i\chi_i(X)\,g_i(k^2_{\perp},x),
\label{chi1}
\eeq
with components
\beq
\chi_i(X)=\Delta_i(X)/\DF(0).
\label{chi_i}
\eeq

Let us again introduce the Fermi averaged  gap-shape function
\beq
\chi_{\Fs}(X)=
\sum_i  \chi_i(X)
g_i(k^2_{\Fs}(X)).
\label{chi_F}
\eeq

This quantity is displayed in Fig.~8 and Fig.~9 at different
$E_0$ and $K_{max}$, correspondingly.

\begin{figure}[h!]\vspace{-2mm}
\begin{center}
\includegraphics[height=80mm,width=100mm]{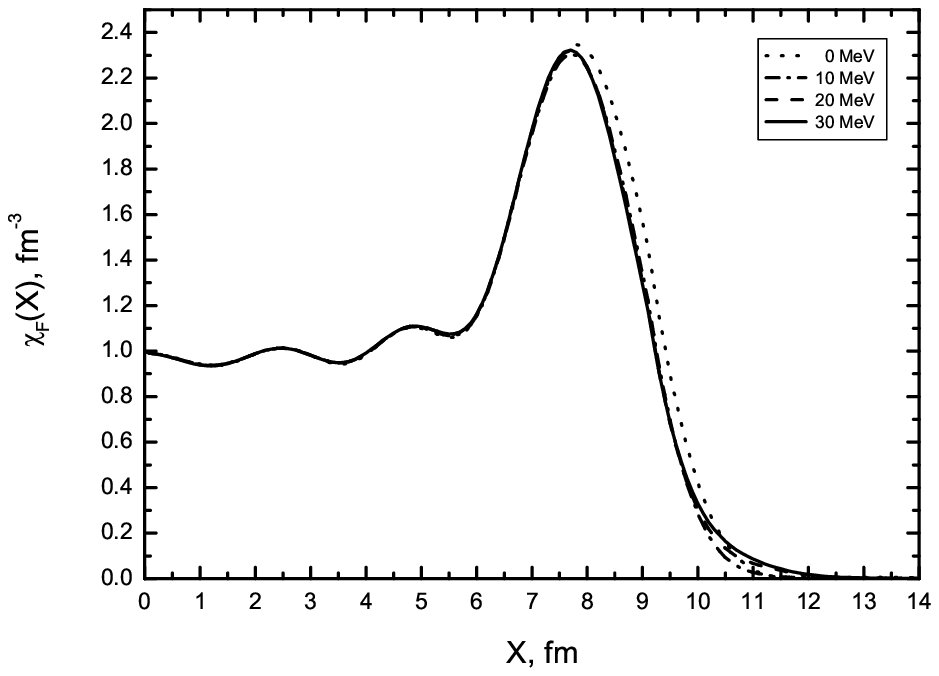}%
\vspace{0mm}
\end{center}
\hskip 1.5 cm
\begin{minipage}[c]{0.8\linewidth}
\caption{ The Fermi averaged gap-shape function  $\chi_{\Fs}(X)$
for different $E_0$.}
\end{minipage}
\end{figure}

\begin{figure}[h!]\vspace{-2mm}
\begin{center}
\includegraphics[height=80mm,width=100mm]{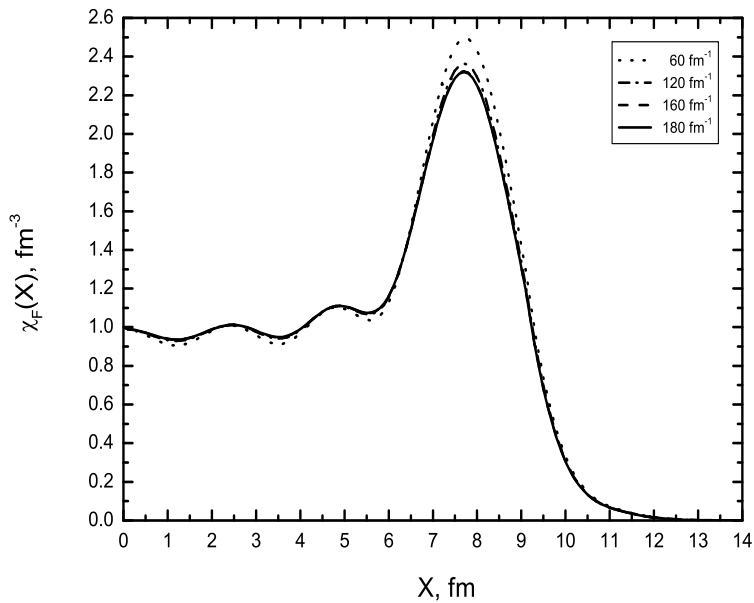}%
\vspace{0mm}
\end{center}
\hskip 1.5 cm
\begin{minipage}[c]{0.8\linewidth}
\caption{ The Fermi averaged gap-shape function  $\chi_{\Fs}(X)$
for different $K_{max}$.}
\end{minipage}
\end{figure}

One can see that the
 gap-shape function is much less sensitive to a variation of these
parameters than the gap itself. In fact, all the sensitivity of the
gap is related to the normalization factor of $\DF(0)$. Thus,
our previous calculations of the  gap-shape function with the use
of $K_{max}{=}60\;$fm$^{-1}$ have sufficiently good accuracy.
Such high sensitivity of the gap value to tiny variations of the
calculation details shows a strong surface-volume interplay. In
other words, it implies a significant proximity effect.

The results of the solution of the gap equation for $L{=}8\;$fm
are presented  in Fig.~10 (the separate components $\Delta_i(X)$)
and in the Table 1 for the diagonal matrix elements
$\Delta_{nn}(k\pr{=}0)$, $n$ corresponding to the states with the
positive parity.

\begin{figure}[h!]\vspace{-2mm}
\begin{center}
\includegraphics[height=80mm,width=100mm]{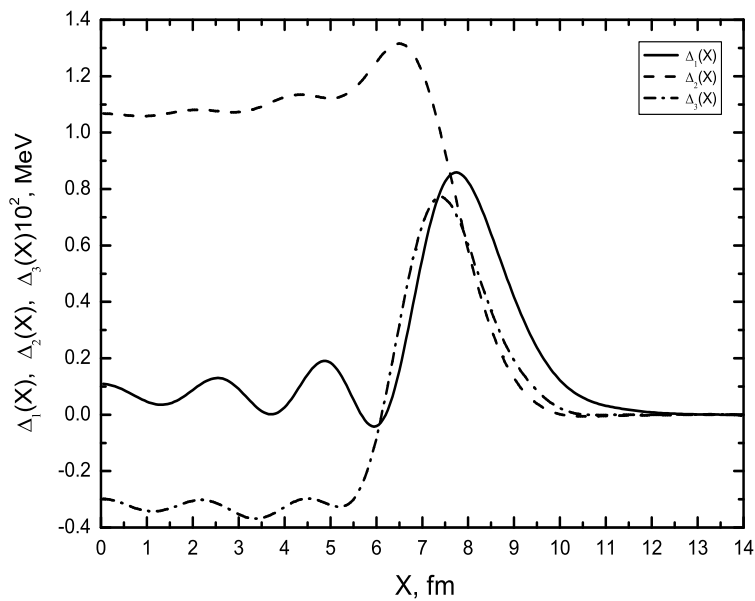}%
\vspace{0mm}
\end{center}
\hskip 1.5 cm
\begin{minipage}[c]{0.8\linewidth}
\caption{ The components of the gap $\Delta_i(X)$.}
\end{minipage}
\end{figure}

 These are the matrix elements which could be
related to those in heavy nuclei over the single-particle
$s$-states. The typical value of the matrix elements $\Delta_{nn}$
is of the order of 1 MeV. The only exception is the one
corresponding to very small energy $\eps_n{=}-1.09\;$MeV. The
reason is quite trivial. Indeed, the wave function has very long
``tail'' outside the slab, therefore the weight of $y_n(x)^2$ in
the integral of $\Delta_{nn}$ is rather small in the region where
the components $\Delta_i(X)$ are large. The average value of the
gap in Table 1 is $\bar{\Delta}{=}0.85\;$MeV (and ${=}1.00\;$MeV,
if to exclude the state with the small energy).
The self-consistent value of the chemical potential
$\mu{=}-7.96\;$MeV is very close to the one, $\mu_0{=}-8\;$MeV,
without pairing. The difference $\delta \mu{=}-0.04\;$MeV is in
agreement with the standard estimate $\delta \mu \sim -
\Delta^2/\eps_{\Fs}$.

\vskip 0.5 cm
\centerline{Table 1. Diagonal matrix elements
$\Delta_{nn}(k_{\pr}{=}0)$ in the slab with the width parameter
$L{=}8\;$fm.}

\begin{tabular}{|c|c|}
\hline
$\eps_n,\;$MeV & $\Delta_{nn},\;$MeV   \\
\hline
  -49.06  & 1.14  \\
  -42.16  & 1.11  \\
  -30.22  & 0.98  \\
  -15.22  & 0.76  \\
   -1.09  & 0.26  \\

\hline
\end{tabular}
\vskip 0.5 cm

Let us now examine dependence of the results on the value of the
width parameter $L$. The Fermi averaged gap functions for
different $L$ are displayed in Fig.~11.

\begin{figure}[h!]\vspace{-2mm}
\begin{center}
\includegraphics[height=80mm,width=100mm]{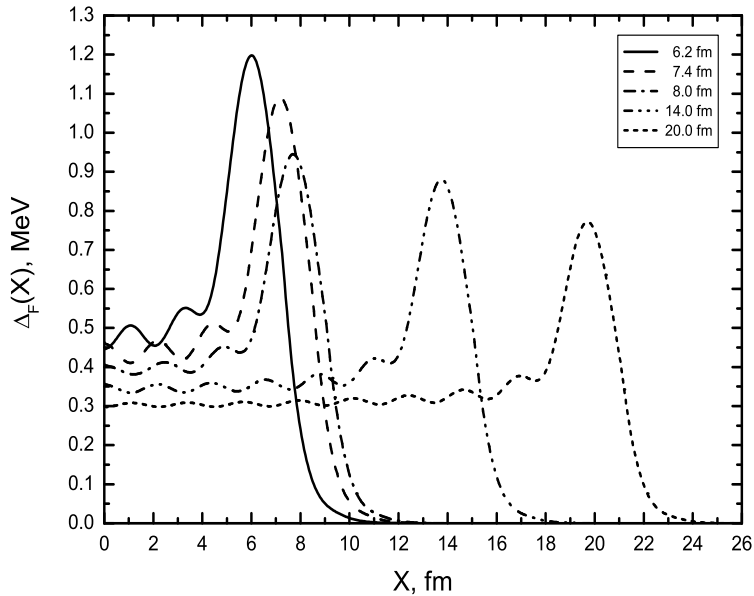}%
\vspace{0mm}
\end{center}
\hskip 1.5 cm
\begin{minipage}[c]{0.8\linewidth}
\caption{ The Fermi averaged gap $\DF(X)$ for different $L$.}
\end{minipage}
\end{figure}

The values of $L{=}6.2\;$fm and $L{=}7.4\;$fm are chosen to
imitate nuclei of the tin and lead regions, correspondingly. The
average diagonal matrix elements of $\Delta(k\pr{=}0)$  are
$\bar{\Delta}{=}1.08\;$MeV for Pb and $\bar{\Delta}{=}1.06\;$MeV
for Sn. In the latter case, again there is a small energy level
with a small value of $\Delta_{nn}$. If it will be excluded from
the analysis, one obtains $\bar{\Delta}{=}1.26\;$MeV. In any case,
these values are rather close to those in atomic nuclei which are
about 1 MeV for the Pb isotopes and about 1.4 MeV, for Sn
isotopes. As it is seen in Fig.~11, with increase of $L$ the
surface maximum of the function $\DF(X)$ becomes lower. At the
same time, the central value $\DF(X{=}0)$ becomes smaller. The
corresponding values are given in Table 2. At the same time, the
ratio $\alpha{}=\DF^{max}/\DF(0)$ depends on $L$ very smoothly.

\vskip 0.5 cm
\centerline{Table 2. Dependence of the cental values $\DF(X{=}0)$ on
the width parameter $L$.}

\begin{tabular}{|c|c|c|c|}
\hline
 $L$, fm & $\DF(X{=}0)\;$,MeV & $\DF^{max}\;$,MeV &$\alpha$\\
\hline
  6.2      & 0.45  & 1.20  & 2.67  \\
  7.4      & 0.46  & 1.09  & 2.37  \\
  8.0      & 0.41  & 0.94  & 2.29  \\
 14.0      & 0.36  & 0.88  & 2.44  \\
 20.0      & 0.30  & 0.77  & 2.57  \\
 $\infty$  & 0.22  &  -    &  -    \\

\hline
\end{tabular}
\vskip 0.5 cm

   The Fermi averaged gap-shape functions at different $L$ are displayed
in Fig.~12. In agreement with Table 2, the values of the surface maximum
are almost $L$-independent. Fig.~12 is in a good agreement with the
analogous one in \cite{Rep} obtained with the help of the KKC method.

\begin{figure}[h!]\vspace{-2mm}
\begin{center}
\includegraphics[height=80mm,width=100mm]{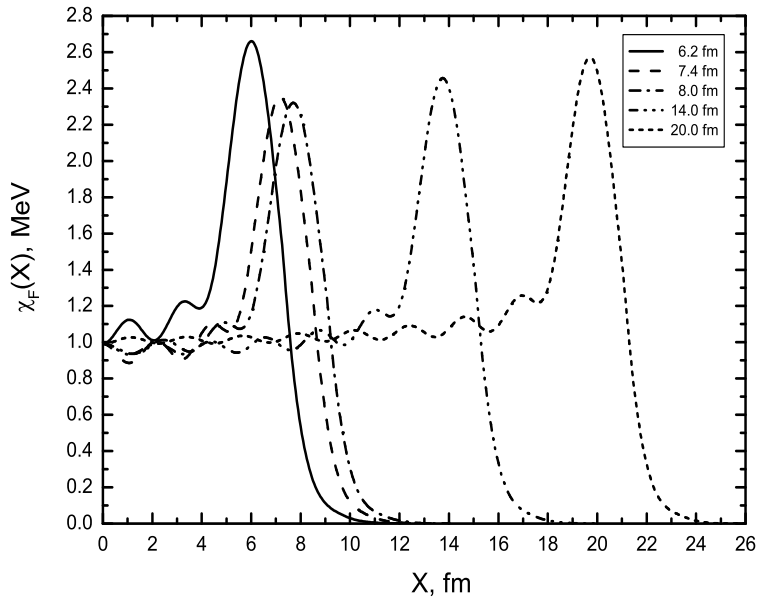}%
\vspace{0mm}
\end{center}
\hskip 1.5 cm
\begin{minipage}[c]{0.8\linewidth}
\caption{ The Fermi averaged gap-shape function  $\chi_{\Fs}(X)$
for different $L$.}
\end{minipage}
\end{figure}

\section{Conclusions}

We solved the gap equation in the $^1S_0$-channel for a nuclear
slab with the separable $3\times 3$ representation
\cite{Par1,Par2} of the Paris potential. A two-step method was
used in which the gap equation is solved in the model space in
terms of the EPI which is found in the complementary subspace. The
model space is chosen sufficiently wide to make it possible to
neglect the pairing effects in the complementary subspace. In this
case, the EPI obeys the Bethe-Goldstone-type equation which was
solved with the help of the so-called local-potential
approximation which, as it was shown in \cite{Rep}, has very high
accuracy. The absolute value of the gap  $\Delta$ turned out to be
very sensitive to the cutoff $K_{max}$ in the momentum space in
the equation for the EPI. It is necessary to take  $K_{max}{=}160
\div 180\;$fm$^{-1}$ to guarantee 1\% accuracy for $\Delta$.

 The gap equation itself is solved directly, without any additional
approximations. The solution shows the surface enhancement of the
gap $\Delta$ which was earlier found \cite{6auth} with an
approximate consideration of the gap equation.
A significant surface-volume interplay was observed also, thereby
a kind of the proximity effect was found. The diagonal matrix
elements of $\Delta$ which could be corresponded to those in heavy
atomic nuclei turned out to be rather close to empirical values.
Of course, there are various corrections to the BCS approximation.
In nuclear matter, they came mainly from the screening of the EPI
due to particle-hole excitations [15--18] and the dispersive
self-energy effects \cite{alb}. In the nuclear slab system under
consideration, just as in real atomic nuclei, they are partially
suppressed due to the enhanced contribution of the surface region.
Evidently, the most important correction to the BCS approximation
is the one due to the surface vibrations \cite{milan}. In our
calculations, the room for all the additional corrections to the
BCS approximation is of order of 20\%. This contradicts  to some
extent to results of Ref.~\cite{Kadm,Kam,milan} where the
microscopic gap equation was solved directly for the spherical
nucleus $^{120}$Sn with Argonne NN-force $v14$, and only about
50\% of the empirical value of $\Delta$ was obtained. The
additional 50\% was explained in \cite{milan} with the surface
vibration contributions. We suppose that the main reason in the
difference of our results of solving the BCS gap equation and
those of Ref.~\cite{milan} is in value of the effective nucleon
mass $m^*$ which was chosen in \cite{milan} as the density
dependent one varying from $m^*{=}0.7m$ inside the nucleus to
$m^*{=}m$ outside. We used the value  of $m^*{=}m$ which is, in
our opinion, closer to the empirical effective mass in atomic
nuclei. As to the surface vibration contribution itself, possibly,
it is partially overestimated in \cite{milan} due to neglecting
the so-called non-pole diagrams. This point is discussed in more
detail in \cite{Rep}. The problem of the consistent account of the
non-pole diagrams in the pairing problem, along the line of
\cite{KhS}, is waiting for its solution.

\section{Acknowlegements}
The authors are highly thankful to S.T.~Belyaev and
S.V.~Tolokonnikov for helpful discussions. This research was
partially supported by the Grant NS-1885.2003.2 of The Russian
Ministry for Science and Education.

\newpage

\end{document}